\documentclass[12pt,preprint]{aastex}
\usepackage{bbm}
\usepackage{mathrsfs}
\usepackage{amssymb,amsmath,float}
\usepackage{rotating}
\usepackage{color}

\newcommand\de{\delta}
\newcommand\ep{\epsilon}

\renewcommand\th{\theta}

\newcommand\om{\omega}

\newcommand\Om{\Omega}

\newcommand\ie{\emph{i.e.}}
\newcommand\eg{\emph{e.g.}}

\newcommand\beq{\begin{equation}}
\newcommand\eeq{\end{equation}}
\newcommand\bea{\begin{eqnarray}}
\newcommand\eea{\end{eqnarray}}
\newcommand\bal{\begin{align}}
\newcommand\eal{\end{align}}

\newcommand\fr{\frac}

\newcommand\ap{\approx}

\newcommand\bB{\bold{B}}

\newcommand\bE{\bold{E}}

\renewcommand\bal{\mbox{\boldmath$\alpha$}}

\begin{document}

\title{The role of Cerenkov radiation in the pressure balance of cool core clusters of galaxies}

\author{Richard Lieu$^1$}

\affil{$^1$Department of Physics, University of Alabama,
Huntsville, AL 35899\\}

\begin{abstract}
Despite the substantial progress made recently in understanding the role of AGN feedback and associated non-thermal effects, the precise mechanism that prevents the core of some clusters of galaxies from collapsing catastrophically by radiative cooling remains unidentified.  In this paper we demonstrate that the evolution of a cluster's cooling core, in terms of its density, temperature, and magnetic field strength, inevitably enables the plasma electrons there to quickly become Cerenkov loss dominated, with emission at the radio frequency of $\lesssim$ 350 Hz, and with a rate considerably exceeding free-free continuum and line emission.  However, the same does not apply to the plasmas at the cluster's outskirts, which lacks such radiation.  Owing to its low frequency, the radiation cannot escape, but because over the relevant scale size of a Cerenkov wavelength the energy of an electron in the gas cannot follow the Boltzmann distribution to the requisite precision to ensure reabsorption always occurs slower than stimulated emission, the emitting gas cools before it reheats.  This leaves behind the radiation itself, trapped by the overlying reflective plasma, yet providing enough pressure to maintain quasi-hydrostatic equilibrium.  The mass condensation then happens by Rayleigh-Taylor instability, at a rate determined by the outermost radius where Cerenkov radiation can occur.  In this way, it is possible to estimate the rate at $\ap 2 M_\odot$~year$^{-1}$, consistent with observational inference.  Thus the process appears to provide a natural solution to the long standing problem of `cooling flow' in clusters; at least it offers another line of defense against cooling and collapse, should gas heating by AGN feedback be inadequate in some clusters.
\end{abstract}

\section{Introduction -- statement of the problem}

It has been known for sometime (\cite{lea73,cow77,fab77}) that the plasma conditions in the center of many clusters of galaxies enable
the region to cool by free-free emission on timescales well within one Hubble time and, consequently, one expects
copious star formation and central mass deposition triggered by this very large cooling rate.  Specifically the cooling time of free-free continuum emission is \beq t_{\rm ff} = 4.24 \times 10^8 \left(\fr{T}{5 \times 10^5~{\rm K}}\right)^{1/2}\left(\fr{n_e}{10^{-2}~{\rm cm}^{-3}}\right)^{-1}~{\rm years}, \label{tff} \eeq and the inclusion of line cooling would shorten this time by $\ap 40$ times for gas at temperatures $\ap 5 \times 10^5$~K and abundance 0.5 solar, see Figure 1 of \cite{sch09} (note the figure assumes full solar abundance).  Such losses are responsible for a theoretical mass deposition rate of
\beq \dot M = \fr{2}{5}\fr{L_X \mu m_p}{kT} = 100 \left(\fr{L_X}{10^{44}~{\rm ergs/s}}\right)\left(\fr{kT}{3~{\rm keV}}\right)^{-1}
\left(\fr{\mu}{0.5}\right)~M_\odot~{\rm year}^{-1}. \label{Mdot} \eeq  where $L_X$ is the X-ray luminosity of the cool core and $kT$ is the temperature.  The actual $\dot M$ inferred from observations at various wavelengths is inconsistent with such a large value, and the ensuing
`cooling flow' problem has lasted for several decades despite numerous attempts in finding a solution, see \eg~the rather detailed analysis of \cite{hud10}.  Although there has recently been a lot of progress made in ascertaining the role of the central AGN in terms of its feedback effects, \eg~\cite{gas13,das16}, the exact physical mechanism that provides the necessary pressure to hold off the inflow and further cooling of such large quantities of matter remains a mystery.

\section{Cerenkov radiation in the magnetized intracluster medium and its competition against free-free emission}

In this paper we suggest that another, hitherto ignored, emission process at play in the plasma under the relevant conditions may hold the key to the cooling flow problem.

It is known that the plasma of the intracluster medium (ICM) has a frozen-in magnetic field, which for the cool cluster cores can be as high as 30-50 $\mu$G, \cite{tay93}, \cite{tay06}, and \cite{fab08}; or as low as a few $\mu G$, \cite{gov06} and \cite{mcn12}.  As conditions of the core we further assume that the central temperature can cool to $kT = 0.05$~keV and the density reaches $n_e = 0.01$~cm$^{-3}$, so that the cooling time is
\beq t_{\rm cool} \ap 10^7~{\rm years} \label{tcool} \eeq as obtained from (\ref{tff}) after taking into account the factor of 40 decrement due to line emission, as mentioned in the text after (\ref{tff}).  Note that in this parameter regime of choice the magnetic pressure equals the gas pressure, while in the outskirts of the cluster where $B \ap 1 \mu$G and $n_e \ap 10^{-3}$~cm$^{-3}$ the gas pressure dominates.  Indeed the importance of magnetic and cosmic ray pressure in cluster cool cores was addressed, see \eg~\cite{lag10} and earlier papers cited therein.

Turning to the focus of this paper, we wish to point out that, in addition to free-free continuum radiation and resonance line emission, electrons in the magnetized plasma of cool cluster cores can also lose energy by Cerenkov radiation (see \cite{mck63,mck67,gin79,mel91}) provided the criterion \beq \cos^2 \theta = \fr{c^2}{\mu^2 (\th) v^2} \label{cercond} \eeq is fulfilled.  In (\ref{cercond}), $\th$ is the angle w.r.t. the local magnetic field $\bB$ at which the radiation is emitted, $v$ is the speed of the electron parallel (or anti-parallel) to $\bB$, and $\mu = \mu (\th)$ is the refractive index of the magneto-ionic medium which depends on the frequency of the emitted wave as well as $\th$ in a complicated way.  Yet, as noted on p600 of the seminal paper of \cite{mck63},  provided the magnetic field is strong enough and $v$ is small, such that the cyclotron frequency $\Om = eB/m_e$ satisfies the inequality \beq \fr{v^2}{c^2}\fr{\om_p^2}{\Om^2} \ll 1 \label{ompom} \eeq
where $\om_p = \sqrt{n_e e^2/(\ep_0 m_e)}$ is the electron plasma frequency,
there exists a principal mode of emission extending from $\om=0$ to a maximum frequency of \beq \om_m = \Om\left(1-\fr{v^2}{c^2}\fr{\om_p^2}{\Om^2}\right), \label{omm} \eeq the relevance of which we shall demonstrate. The radiation is fairly isotropic if $v \ll c$.  Note also in passing that there is a second mode of emission (both modes are consequences of the Alfv\'en-Whistler approximation limit), by which the electron loses energy at a relatively negligible rate, and will not be discussed here.

Before continuing further, one should beware the caveats.  The underlying assumptions of the last three equations are three-fold.  First, the wavelength $\lambda$ of the emitted radiation far exceeds the size of any homogeneous subregion of the ICM within which $\bB$ and $\om_p$ do not vary spatially. Since the typical scale size of field smoothness is of order the gyroradius of a proton in the same field $\bB$, this size is indeed $\ll \lambda$.  Second, the frequency of collisions of plasma particles is small w.r.t. $\Om$ and $\om_p$ (the approximation of a collisionless plasma); as we shall show below in (\ref{tcol}), this too is always the case for the ICM.  Third, the results apply to the regime of a cold plasma,\ie~if thermal motions are included there will be relatively small corrections which we neglected.

Another interesting feature about (\ref{ompom}) is that although the criterion holds for the cooling core parameters, it does not for the outskirts of a cluster where $v \ap 8 \times 10^8$~cm~s$^{-1}$ ($kT \ap 5$~keV), $n_e \ap 10^{-3}$~cm$^{-3}$, and $B \ap 1 \mu$G.  The plasma is essentially unmagnetized and cannot support Cerenkov radiation.

Hence Cerenkov losses are significant {\it only} in the core.  The spectral emissivity of this mode is \beq \fr{dI}{d\om} = \fr{e^2\mu_0}{4\pi}\fr{c^2}{v}\fr{1}{\om_p^2} \om(\Om^2 - \om^2),  \label{dIdw} \eeq see eqs. (6.2) of \cite{mck63} and (33) of \cite{mck67}.  The total power emitted into the frequency range $0<\om < \Om$  is \beq I = 1.16 \times 10^{-12}\left(\fr{v}{2.37 \times 10^8~{\rm cm~s}^{-1}}\right)^{-1}\left(\fr{B}{20 \mu {\rm G}}\right)^4
\left(\fr{n_e}{10^{-2}~{\rm cm}^{-3}}\right)^{-1}~{\rm eV~s}^{-1}, \label{P} \eeq where
\beq \Om = \fr{eB}{m_e} = 351.2 \left(\fr{B}{20~\mu {\rm G}}\right)~{\rm Hz}, \label{omc} \eeq $v$ is the projected average {\it speed} along the $+\bB$ or $-\bB$ direction of the particles in an isotropic thermal gas, {\it viz.} \beq v = \sqrt{\fr{2kT}{\pi m_e}} \ap 2.37 \times 10^8~\left(\fr{kT}{0.05~{\rm keV}}\right)^{1/2}~{\rm cm~s}^{-1}. \label{v} \eeq  In enlisting only the parallel component of ${\bf v}$ for the sake of simplicity, one underestimated $dI/d\om$ somewhat.  The power spectrum associated with the full helical motion of the electron may be found in \cite{mck67}.

\section{Comparison to conventional cooling rate; re-absorption}

\begin{figure}
\begin{center}
\includegraphics[angle=0,width=7in]{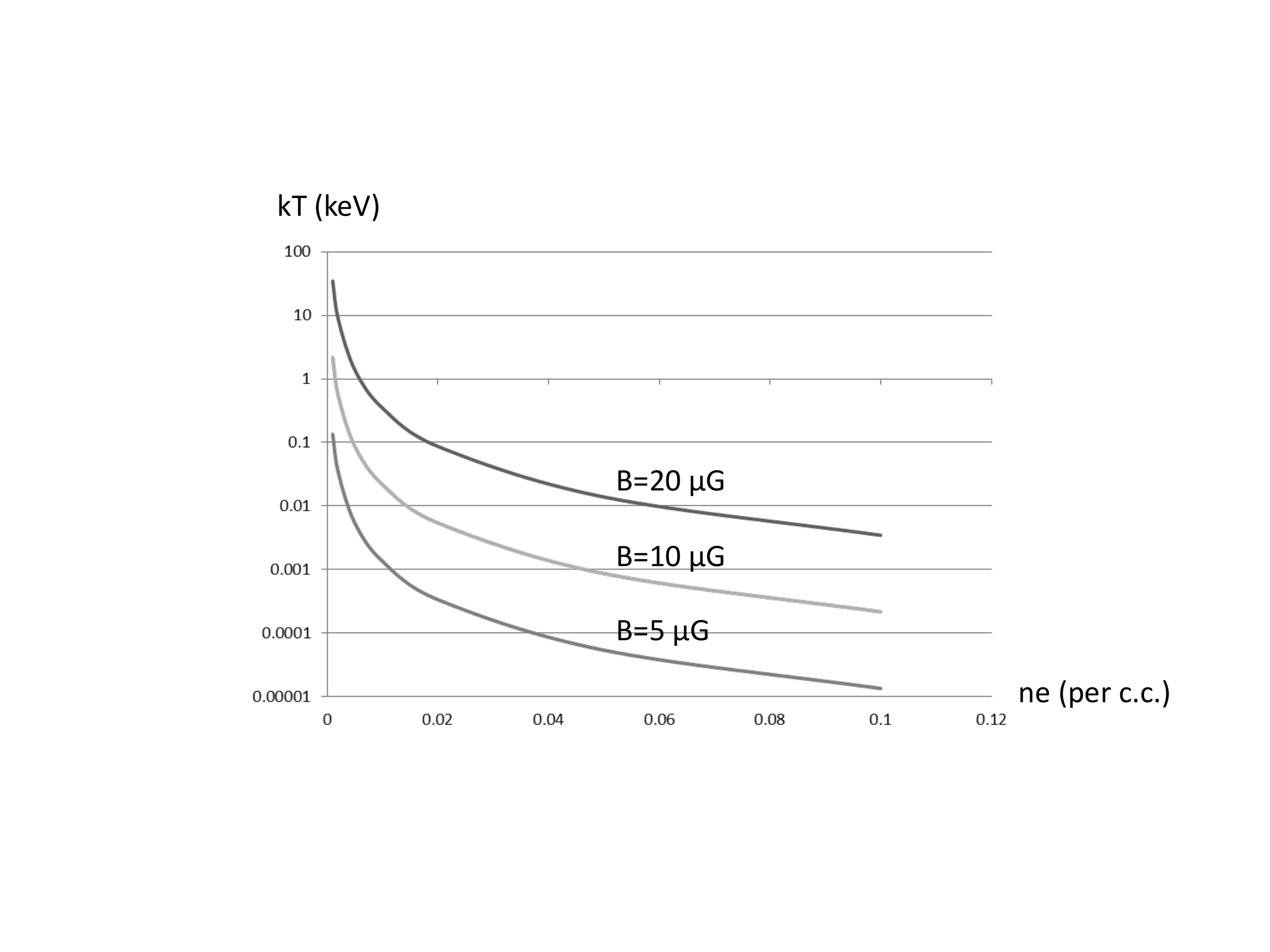}
\end{center}
\caption{Critical values of ICM density and temperature, below which Cerenkov radiation cools the gas faster than X-ray emission.  All points on the curves satisfy (\ref{ompom}).  For each ICM magnetic field, points lying below the corresponding curve fall under the regime of Cerenkov loss domination.}

\label{cartoon}
\end{figure}

In (\ref{P}) the reader is alerted to the particle density and speed dependence of the Cerenkov intensity; they are counterintuitive because one is used to thinking about a conventional emissivity like free-free and resonant transition, which {\it increases} with $n_e$ and $v$.  In fact, from (\ref{P}) one can readily estimate the lifetime of a $kT = 0.05$~keV emitting electron, as \beq t_{\rm Cerenkov} \ap 1.44 \times 10^6 \left(\fr{kT}{0.05~{\rm keV}}\right)^{3/2} \left(\fr{B}{20~\mu {\rm G}}\right)^{-4} \left(\fr{n_e}{10^{-2}~{\rm cm}^{-3}}\right)~{\rm years}, \label{tC} \eeq which is considerably shorter than the lifetime against free-free and line cooling, (\ref{tcool}).  It is also shorter than the lifetime against cooling by cyclotron radiation \beq t_{\rm cyclotron} \ap 4.1 \times 10^{11} \left(\fr{kT}{0.05~{\rm keV}}\right)^{-1} \left(\fr{B}{20~\mu {\rm G}}\right)^{-2}~{\rm years} \label{tcyclo} \eeq by 6 orders of magnitude.

Thus it is clear that cool cluster cores possess certain unique characteristics not commonly shared by other astrophysical environments, {\it viz.} they are venues where Cerenkov radiation can take place on large, 10 -- 100 kpc, scales.  In Figure 1 we plot the critical temperature and density values for which $t_{\rm Cerenkov} = t_{\rm cool}$ at two values of $B$, where $t_{\rm cool}$ is the X-ray cooling time, assumed to be shorter than (\ref{tff}) by 40 times because of line emission.  One can see that even for fields as weak as a few $\mu G$ the gas temperature cannot cool below 0.5 keV at a density of $n_e = 0.01$~cm$^{-3}$.  It should be emphasized, however, that Cerenkov radiation can prevent cooling flow only when the parameters of the {\it entire} cluster core place it below the corresponding critical curve.   Nevertheless, the fact that for every parcel of cooling gas in the ICM this must happen at some stage can also be understood from general physical arguments.  If the ICM cools and condenses, $B \propto 1/r^2$ by flux conservation and $n_e \propto 1/r^3$ by mass conservation.  The quantity on the left side of (\ref{ompom}) which is proportional to $n_e/B^2 \propto r$, would then decrease with $r$ to quickly evolve the gas to satisfy this inequality of the Cerenkov condition; and the lifetime against Cerenkov radiation, (\ref{tC}), would decrease even faster, as $n_e/B^4 \propto r^5$ to become less than the X-ray cooling time.   Indeed, \cite{sok90} provided a formula for the radius at which the magnetic pressure of a cooling flow cluster will inevitably dominate the gas pressure.  Thus the purpose of this paper is to point out that even in the absence of heating by AGN feedback, there exists another mechanism that may offer a second line of defense against cooling flow.

Apart from the existence of this radiation, there is the question of its fate after emission.  Unlike the X-ray, EUV, and visible photons, which can usually stream out of the optically thin core, Cerenkov radiation consists of  very low frequency radio waves, (\ref{omc}), which can only propagate in the magnetoionic medium of the core at fixed angles to the local magnetic field, \cite{mck63}.  Since the field lines are unlikely to be smooth and radially directed on 10 -- 100 kpc scales, the radiation will not escape the core region; even if they do, the outer ICM region of relatively unmagnetized plasmas has a plasma frequency far in excess of (\ref{omc}), \ie~it will reflect the radiation back to the center.

Can the Cerenkov photons be re-absorbed by the core and reheat the gas there?  Since they are unable to escape, the answer is: yes, eventually.  In fact, even for an isotropic unmagnetized plasma, provided the temperature is finite, longitudinal plasma waves can propagate in lieu of electromagnetic waves, and the re-absorption of such waves is the same process as Landau damping, see p142 and 260 of \cite{gin79}.  Yet there is another related point that must also be taken into account.  Absorption is effective only if its rate is higher than stimulated emission.  For a perfectly Maxwellian thermal gas, the rate of absorption is indeed higher than stimulated emission, fractionally by the amount $1 - \exp [-\hbar\om/(kT)] \ap \hbar\om/(kT)$, see \eg~\cite{bek66}.

Since the maximum emitted frequency $\om_m$ corresponds to $\hbar\om/(kT) \ap 4.4 \times 10^{-15}$, the Maxwell-Boltzmann energy distribution $f(\ep) \propto \exp [-\ep/(kT)]$ has to be extremely stable to ensure that absorption is always higher, everywhere in the gas. In fact, it reasonable to assert that if the emitted Cerenkov radiation is re-absorbed and cannot as a result exert any formidable pressure on the gas, the mean energy fluctuation $\de\ep/\ep$ of a particle in any volume of gas of size $\ap$ one radiation wavelength $\lambda$ has to be $\ll \hbar\om/(kT)$.  Given that $\lambda^3 \ap 1.6 \times 10^{26}$~cm$^{3}$, for $n_e \ap 0.01$~cm$^{-3}$ one has $N \ap 10^{24}$ particles, so that $\de\ep/\ep \ap 1/\sqrt{N} \ap 10^{-12} \gg \hbar\om/(kT)$, the gas is not stable enough to ensure absorption prevails.  One must therefore assume, due to the much larger random fluctuations in the particle energy, there is population inversion among half of the electrons, \ie~at least a significant fraction of the emitted Cerenkov radiation can survive absorption\footnote{One can also arrive at the same conclusion by applying a {\it reducio ad absurdum} argument.  If the electron energy distribution follows strictly the Boltzmann factor $\exp [-\ep/(kT)]$ even at the resolution of $\de\ep = \hbar\om\lesssim 2 \times 10^{-13}$~eV, the mean energy per particle in the volume of $\lambda^3$ will {\it inevitably} have to satisfy the inequality $\de\ep/\ep \ll 1/\sqrt{N}$, in violation of a standard statistical mechanics result that may be found in textbooks like \cite{man91}.} .  Of course, the absorbing fraction of the electron population cannot reheat itself to the extent of taking back from the radiation the energy of the remaining (unabsorbed) fraction.  The net result is continuous cooling of the electrons, as they keep emitting more net radiation than they can absorb, until all their energy is converted to Cerenkov radiation.

Can the gas be reheated by electron scattering with the Cerenkov photons?  Since the photon energy $\hbar\om$ is smaller than the electron there can be no energy transfer from the former to the latter.  The only possibility is the non-linear Compton process, but this requires the incident wave to have a sufficiently large amplitude, satisfying the condition $\Gamma = eE/(m_e c \om) \gg 1$ with $\tfrac{1}{2} \ep_0 \bE^2 = 3\eta n_e kT$, see \eg~p266 of \cite{sok86}.  The reader can readily verify that this $\Gamma$ parameter of the Cerenkov radiation here is not large enough; in fact \beq \Gamma = 0.123 \left(\fr{\eta}{0.15}\right)^{1/2} \left(\fr{kT}{0.05~{\rm keV}}\right)^{1/2} \left(\fr{\om}{350~{\rm Hz}}\right)^{-1}. \label{Gamma} \eeq  In any case, the mean free path of electron scattering is much larger than the size of the cluster core for there to be meaningful acceleration of gas electrons by this mechanism.

\section{Pressure of Cerenkov radiation; quasi-hydrostatic equilibrium; limit to the mass deposition rate}

For reasons explained in the last section, the energy of the electrons in the thermal gas of a cluster's cool core is emitted as Cerenkov radiation at a sufficiently early stage, {\it before} further cooling by X-ray and EUV emission can act.  In this way the entire gas will lose its energy, because the ions will follow suit: the ion-electron collision timescale \beq t_{\rm collision} = 23.6 \left(\fr{kT}{0.05~{\rm keV}}\right)^{3/2} \left(\fr{n_e}{10^{-2}~{\rm cm}^{-3}}\right)^{-1} \left(\fr{\ln\Lambda}{17}\right)^{-1}~{\rm years}, \label{tcol} \eeq where $\Lambda = b_{\rm max}/b_{\rm min}$ and $b$ is impact parameter, is instantaneously short.  Thus, by the time all the gas in the core has cooled and condensed, the Cerenkov radiation holds essentially the same reservoir of energy density as the gas initially.  And because the radiation is trapped by the overlying plasma, it cannot escape.

If the pre-cooled gas was in hydrostatic equilibrium in the cluster potential well, the radiation trapped in the core by the overlying layers of hot gas will, after some configurational adjustment of scale heights (because radiation having the same energy density as the gas has only half the pressure) to be worked out later in a more detailed paper, prevent further episodes of central condensation and accelerated cooling.   The purpose of this Letter is to highlight the importance of a much faster electron loss mechanism than any of the ones hitherto known, by enlisting a set of parameters representative of the cooling gas at some radius within the core of a cluster.  Of course, we merely provided a working example, as the gas parameters depends on radius, but nevertheless they serve to demonstrate the existence of an entire central region of Cerenkov radiation {\it pressure} domination, defined by a boundary radius beyond which no emission can take place because the inequality (\ref{ompom}) is no longer satisfied.

The exact value of such a boundary radius may vary from cluster to cluster, and is given by the point on Figure 1 where the gas parameters first find themselves below the curve corresponding to the ICM core magnetic field (note the argument given after (\ref{cyclo}), which proves that as the gas cools the density and temperature {\it must} quickly place it below the curve).   Thus \eg~for an ICM with $kT \ap$~0.05 keV, $n_e \ap 2 \times 10^{-3}$~cm$^{-3}$, and
$B \ap 4 \mu$G (these parameters marginally satisfy (\ref{ompom}),
$\dot M \ap 2~M_\odot$~year$^{-1}$, which is not far away from the observationally inferred values and limits.  Owing to the central (Cerenkov) radiation pressure, the actual mass deposition can only occur sporadically by Rayleigh-Taylor instability, when the overlying matter piles up enough to break through.  Of course, the above parameters are not the only possible combination, \ie~there can be a range of allowable $\dot M$, though it is clear the rate cannot be as large as (\ref{Mdot}).  At least one may conclude that Cerenkov radiation should not henceforth be completely omitted from consideration when modeling of cool core clusters, it might even be a vital ingredient to be included.

The author thanks Don Melrose and Ming Sun for helpful discussions.

\end{document}